\documentclass[a4paper,conference]{IEEEtran}
\pdfoutput=1
\usepackage{fancyhdr}
\usepackage{lastpage}
\usepackage[ruled,vlined,boxed]{algorithm2e}
\usepackage{tikz}  
\usetikzlibrary{shapes,arrows}
\usepackage{amsmath}
\usepackage{amsthm}
\usepackage{amssymb}
\usepackage{blindtext}
\usepackage{algpseudocode}
\usepackage{algorithm2e}
\usepackage{algcompatible}
\usepackage[english]{babel}
\usepackage{siunitx}
\usepackage{graphicx}
\renewcommand{\figurename}{Figure.}
\usepackage{float}
\usepackage[nolist]{acronym}
\usepackage{comment}
\usepackage{soul}
 \begin{acronym}
 \acro {dl} [DL] {Deep Learning}
 \acro{los} [LOS] {Line-of-Sight}
 \acro {nlos} [NLOS]{Non-Line-of-Sight}
 \acro{ch} [CH]{Coverage Hole}
 \acro{uav}[UAV]{Unmanned Aerial Vehicle}
 \acro{agv}[AGV]{Automated Guided Vehicle}
 \acro{qos}[QoS]{Quality of Service}
 \acro{csi}[CSI]{Channel State Information}
 \acro{UE}[UE] {User Equipment} 
 \acro{cfr}[CFRs]{Channel frequency responses} 
 \acro{rss}[RSS] {Received Signal Strength}
 \acro{gbsm}[GBSM]{Geometry-Based Stochastic channel Model}
 \acro{dl}[DL]{Downlink}
 \acro{ul}[UL]{Uplink}
 \acro{3gpp}[3GPP]{3rd Generation Partnership Project}
 \acro{los}[LoS]{Line-of-Sight}
 \acro{ml}[ML]{Machine Learning}
 \acro{lstm}[LSTM]{Long-Short Term Memory}
 \acro{tdd}[TDD]{Time-Division Duplex}
 \acro{ofdm}[OFDM]{Orthogonal Frequency Division Multiplexing}
 \acro{ls}[LS]{Least-Squares}
 \acro{snr}[SNR]{Signal-To-Noise Ratio}
 \acro{mmse} [MMSE]{Minimum Mean Square Error}
 \acro{nmse} [NMSE]{Normalized Mean Square Error}
 \acro{mse} [MSE]{ Mean Square Error}
  \acro{qam} [QAM]{Quadrature Amplitude Modulation}
  
  \acro{ssnr} [SSNR]{Symbol-Signal-to-Noise Ratio}
  \acro{snr} [SNR]{Signal-to-Noise Ratio}
\end{acronym} 
\begin{document}
\title{Adaptive Prediction Approach for 3D
Geometry-based communication}
\author{\IEEEauthorblockN{Mervat Zarour\IEEEauthorrefmark{1} Qiuheng Zhou\IEEEauthorrefmark{1} Sergiy Melnyk\IEEEauthorrefmark{1} and Hans D. Schotten\IEEEauthorrefmark{1}\IEEEauthorrefmark{2}}
\IEEEauthorblockA{\IEEEauthorrefmark{1}Intelligent Networks Research Group,
German Research Center for Artificial Intelligence, \\Kaiserslautern, Germany\\\{mervat.zarour, qiuheng.zhou, sergiy.melnyk, hans\_dieter.schotten\}@dfki.de}
\IEEEauthorrefmark{2}Department of Wireless Communication and Navigation, University of Kaiserslautern-Landau, \\Kaiserslautern, Germany \\schotten@rptu.de}
\twocolumn[
  \begin{@twocolumnfalse}
    \maketitle
    \vspace{-1em}
    \centering
    \noindent\textbf{IEEE\textcopyright\ This work has been submitted to the IEEE conference for possible publication. 
    Copyright may be transferred without notice, after which this version may no longer be accessible.}
    \vspace{1em}  
  \end{@twocolumnfalse}
]
\begin{abstract}
This paper addresses the challenges of mobile user requirements in shadowing and multi-fading environments, focusing on the \ac{dl} radio node selection based on \ac{ul} channel estimation. One of the key issues tackled in this research is the prediction performance in scenarios where estimated channels are integrated. An adaptive deep learning approach is proposed to improve performance, offering a compelling alternative to traditional interpolation techniques for air-to-ground link selection on demand. Moreover, our study considers a 3D channel model, which provides a more realistic and accurate representation than 2D models, particularly in the context of 3D network node distributions. This consideration becomes crucial in addressing the complex multipath fading effects within geometric stochastic 3D 3GPP channel models in urban environments. Furthermore, our research emphasises the need for adaptive prediction mechanisms that carefully balance the trade-offs between \ac{dl} forecasted frequency response accuracy and the complexity requirements associated with estimation and prediction. This paper contributes to advancing 3D radio resource management by addressing these challenges, enabling more efficient and reliable communication for energy-constrained flying network nodes in dynamic environments.
\end{abstract}
\section{Introduction}
Wireless communication networks are steadily gaining their popularity in all aspects of business and social life. Advances in wireless technologies, such as 5G and the upcoming 6G, promise to deliver faster data rates or reduced latency. Among others, these improvements will open up new use cases in the field of real-time applications, e.\,g: autonomous vehicles, wireless robotics control, and more. However, the critical challenge on the way to their realisation remains the reliability of the communication links. In outdoor scenarios, 2D terrestrial networks may be enhanced by the utilisation of the third coordinate using airborne wireless nodes.
3D radio resource nodes enable stable and highly reliable links to users, thanks to their high position flexibility within a 3D space. This capability addresses wireless signal propagation, blockage, reflection, shadowing, and the multipath effect. Compared to wireless links established through fixed terrestrial base stations, 3D radio resource nodes significantly improve wireless link quality. 
The flight nodes can serve as base stations or relays and has high flexibility to improve the connectivity link quality with optimal 3D trajectory planning or altitude change, unlike the fixed terrestrial radio access node with a low \ac{los} probability \cite{coexitenceuav}. In 3D networking, ongoing research explores the advantages of employing over-the-air network nodes to enhance \ac{qos} while minimising transmission time due to energy limitation~\cite{UAVaidedcommunication}.  In the mobility control use case, through consecutive control input via a wireless channel, the radio access node selection can support a highly stable and reliable link considering the ground-to-ground link prediction with high accuracy and energy constraint of the flight nodes \cite{5GWithUAVs}. The author optimised the wireless service to the ground user using optimal cell partitioning, considering the average hover times of the \acp{uav} in~\cite{hoverduration}.

Whereas utilising flight nodes is advantageous for maintaining a reliable communication link, the \acp{uav} suffers from limited energy storage capability.  To increase the energy efficiency of the flight nodes, we propose a 3D radio resource management system, which optimises the utilisation of power-demanding air-to-ground links based on \ac{qos} requirements. Thus, the ground-based users should preferably maintain the ground-to-ground link. In case of the ground link degradation due to i.\,e. fading, shadowing, or other effects, the air-to-ground link selection is performed preventing the violation of \ac{qos} requirements.
Establishing an accurate 3D radio resource management system, relying on precise decision-making for \ac{dl} adaptive transmission over-the-air links instead of ground-to-ground links is imperative~\cite{adaptivetrasmissionairnode}. So, the DL selection cannot be accurate when it is based on outdated channel information using uplink channel estimates due to the rapid variation in channel conditions, especially in dynamic environments such as urban areas with mobile \ac{UE}.

\section{System model}\label{section2}  
\begin{figure}
    \centering
    \includegraphics[width=0.45\textwidth]{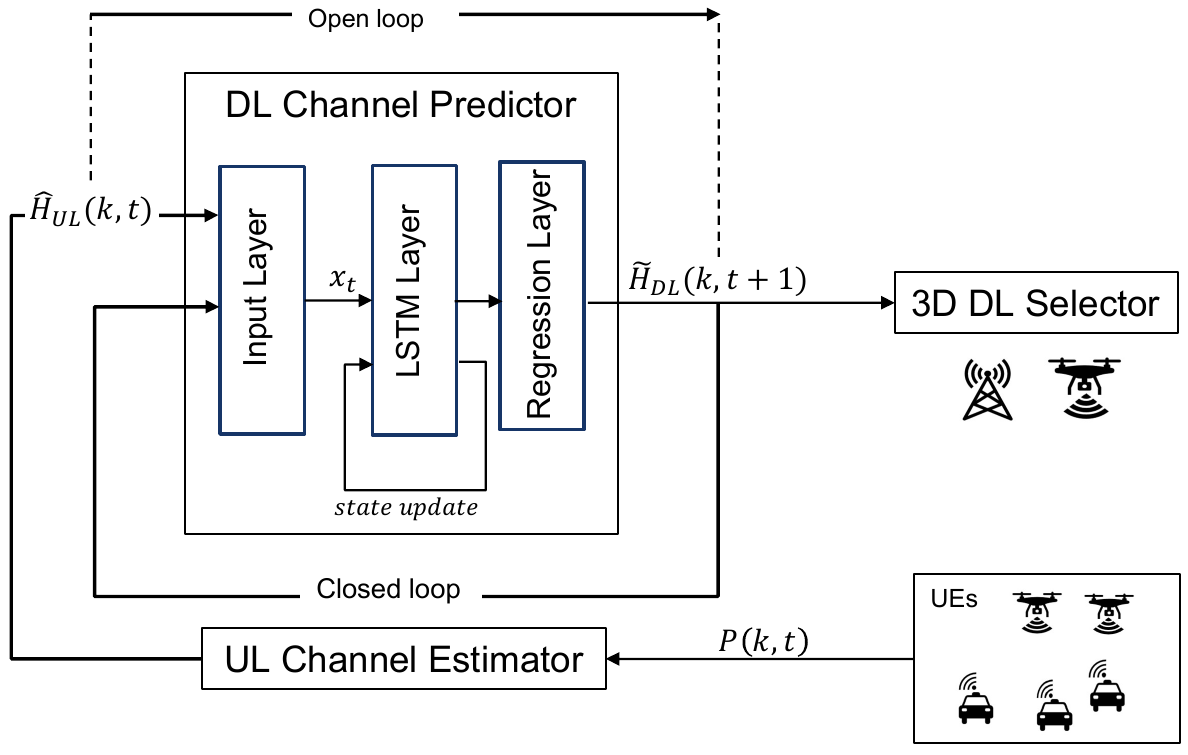}
    \caption{System model}
    \label{fig:1}
\end{figure}
In an Urban environment, channel state information (CSI) changes dynamically, necessitating \ac{dl} selection for optimising 3D radio resources and ensuring radio link quality through a closed-loop continuous monitoring and optimisation process. Air-to-ground \ac{dl} usage occurs on demand to save flight node radio resources and guarantee the provided user link reliability, guided by decisions from the ground-to-ground \ac{dl} selector. The proposed framework, illustrated in figure~\ref{fig:1}, deals with radio channel propagation problems considering the 3D channel modelling. The wireless channel between transmitter and mobile UEs in 3D dimension, exhibits both slow fading and fast fading due to shadowing, multipath and doppler shift effects. This considers the 3D positions of both the receiver and transmitter and signal propagation in azimuth and elevation domain, providing a highly accurate description of signal propagation in a 3D network.

At the transmitter, the received pilot signal $P(k,t)$ in frequency block type serves to estimate the \ac{ul} channel frequency response $\hat{H}_{UL}(k)$ for each subcarrier $k$ at the time $t$. The channel frequency responses ${H}_{DL}(k)$ for the upcoming \ac{dl} are then interpolated linearly based on the estimated values.
Our comparison explores the use of classical linear interpolation alongside deep learning-based \ac{dl} channel prediction $\tilde{H}(k,t)$, built upon the estimation with two predictor structures.

In an open-loop fashion, the predictor updates the internal \ac{lstm} state based on the previously estimated value at each estimated subcarrier at time $t$. In the closed-loop scenario, the predictor forecasts the upcoming channel frequency response at each subcarrier based on the previous predicted value.

The proposed platform optimises the \ac{dl} selector data using different channel estimation methods to improve the accuracy of the \ac{ul} channel frequency responses. This is achieved using an adaptive predictor, which fits our computational demand and requirements and weighs the accuracy of the predicted \ac{dl} data.

The data exchange between the transmitter and the receiver is performed in a framed manner. Each frame has a time interval of $dt$ for the \ac{ul} and \ac{dl} channel via the \ac{tdd} technique. 

The estimated \ac{ul} channel state information $\hat{H}$ can be outdated due to the fading effect when used at the selector input for the upcoming \ac{dl} time slot. Then, the Link selector can serve the user with an air-ground \ac{dl} when the ground-ground \ac{dl} no longer meets the requirements.
 
The proposed channel model, estimator, and predictor are described in the following \ref{subsection2-1},\ref{subsection2-2},\ref{subsection2-3}.
\subsection{ 3GPP 3D channel model} \label{subsection2-1}
The 3GPP 3D channel modelling represents wireless channel propagation in a mobile user-scattering environment. We propose an urban microcell environment where the positions of the scatterers are not precisely determined but are characterised using stochastic channel parameters. We proposed a 3GPP 3D channel model to simulate the 3D ground-to-ground channel between a single transmitter and a mobile receiver\cite{3dsim}.

Considering several key factors with its cross-correlation matrix, including shadowing effects, delay spread, azimuth, and zenith angle spread in both the arrival and departure directions. The number of considered channel taps is $L$. The channel response $h_l$ for the channel tap $l$ is represented as
\begin{equation}
h_l= \sqrt{ (P_l)}.{(\gamma_l^a,\theta_l^a)}^T.\mathcal{F}.(\gamma_l^d,\theta_l^d).\exp^{-j.2\pi.f.d}
\end{equation}

Where $P_l$ is the power of the $l^{th}$ path, and $d$ distance between $(x,y,z)$ transmitter and receiver positions, $f $ frequency, $(\gamma_l^a,\theta_l^a)$ arrival angles, and $(\gamma_l^d,\theta_l^d)$departure angles are determined via stochastic distributions based on measurements as detailed in ~\cite{3gpp}. Where uncorrelated large scale parameter described as a vector of stochastic variables $X\sim \mathcal{N}(0,1) $: delay spread $X^{DS}$  and angel spread of departure ${X^{\theta}}^d, {X^{\gamma}}^d $ and of arrive ${X^{\theta}}^a, {X^{\gamma}}^a $. The shadowing effect via  $X^{SF}$  and ricean $K$ factor via $X^{KF}$. The spatially exponential correlated vector X with de-correlation distance $d_{dec}$ means that the close-set UEs within $d_{dec}$ experience the same attention or path delay due to, e.g., common obstacle position. The path gain $P_l$ depends on the 3D distance, frequency, and the cross-correlated $X_{cor}= M.X$ where M is the cross-correlation matrix from ~\cite{3gpp}. $\mathcal{F}$ is the coupling matrix to capture the polarisation change between the receiver and transmitter.
\subsection{Uplink channel estimator} \label{subsection2-2}
Channel reciprocity in \ac{tdd} \ac{ofdm} system is precious in the adaptation of \ac{dl} channel transmission. In the \ac{ul} channel propagation duration, channel estimation is established to have an accurate channel response for each sub-carrier $k$ in \ac{ofdm} system. Then, it can be based on channel reciprocity, assuming the channel response in the \ac{dl} duration can interpolate it for the upcoming time slot~\cite{estint}. Some decision-making for resource allocation, e.g. link selection, can be planning. The transmission adaption in \ac{dl} gain depends on the channel variation behaviour and estimation accuracy.
The channel frequency response is represented for each \ac{ofdm} sub-carrier. The $H(k)$ is the true channel frequency response for sub-carrier $k$.

\ac{ls} channel estimation estimates the frequency response of the wireless channel based on minimising the sum of squared errors between the received pilot symbols and the at-receiver known transmitted pilot symbols $P(k)$ for subcarrier $k$. It can be expressed as follows~\cite{LSes}:
 \begin{subequations}
 \begin{align}
     P_r(k)=P(k)\cdot H_{\text{UP} } + \sigma_n 
     \end{align}
     \begin{align}
    \hat{H}_{LS}(k)= P(k)^{-1} \cdot P_r(k)
    \end{align}
\end{subequations}
\ac{ls}\ac{mmse} channel estimation finds the estimated channel response $\hat{H}(k)$ to minimise the mean-squared error of the estimation with considering the statistical property of the noise and the channel, so the \ac{ls}\ac{mmse} estimation for the $k^th$ subcarrier is $\hat{H}_{\text{LSMMSE}} $ at the pilot ${P(k)}$, can be as follows\cite{MMSEes}:
 
 \begin{subequations}
    \begin{align}
        R_{hh} = \mathbb{E} \{{H_{\text{UP} }\cdot H_{\text{UP}}^H }\}
    \end{align}
    \begin{align}
        \mathcal{W} = \frac{\beta \cdot {\sigma_n}^2}{\mathbb{E}\{P(k)^2\}}  
        \label{eq:3b}
   \end{align}
     \begin{align}
       \hat{H}_{\text{LSMMSE}}(k) = R_{hh} \cdot (R_{hh} + \mathcal{W} \cdot {I})^{-1} \cdot \hat{H}(k)_{\text{LS}}
          \end{align}
    \end{subequations}
 Where the $\mathbb{E}$ is the expectation value, $R_{hh}$ is the channel auto-correlation matrix, $\sigma_{n}$ the noise variation, and $\beta$ is constance dependent on signal constellation, in table ~\ref{tab:1}.
\subsection{Downlink channel predictor} \label{subsection2-3}
Based on the estimated \ac{ul} channel response frequency $\hat{H}(k)$, can obtain an accurate, valuable channel response for adaptive transmission in the \ac{dl} direction using a recurrent neural network, \ac{lstm} approach.

The accuracy of \ac{lstm} prediction depends on the purity of the input data to be used to update each memory cell $c^i(t)$ and the hidden state $h^i(t)$ for cell $i$ at time $t$ in the \ac{lstm} layer. Specifically, two approaches are to be established: First, the \ac{lstm} layer uses the current value from the estimator as input to update the memory cell and hidden state and then predicts the next value in the sequence.

Second, the \ac{lstm} predictor alternatively uses its own previous predicted value $\tilde{H}(k,t-1)$ as input to predict the channel response value for the upcoming time slot. So the \ac{lstm} cell has several components at each time $t$, including input gates $ig_t$, forget gates $fg_t$, output gates $og_t$, and the cell state $c_t$. These variables contribute to building and updating the hidden state for each \ac{lstm} cell $i$.  So the \ac{lstm} recurrent layer output can be compressed as follows:
Let the $x_t$ be the \ac{lstm} layer input from the sequence input layer.
\begin{subequations} 
\begin{align}
  ig^i_t = \sigma(W_{xi}x_t + W_{hi}h_{t-1} + b_i)  
\end{align}
\begin{align}
fg^i_t = \sigma(W_{xf}x_t + W_{hf}h_{t-1} + b_f)  
\end{align}
\begin{align}
\tilde{c}^i_t = \tanh(W_{xg}x_t + W_{hg}h_{t-1} + b_g) 
\end{align}
\begin{align}
og^i_t = \sigma(W_{xo}x_t + W_{ho}h_{t-1} + b_o)  
\end{align}
\begin{align}
c^i_t = f_t \odot c_{t-1} + i_t \odot \tilde{c}^i_t
\end{align}
\begin{align}
h^i_t = o_t \odot \tanh(c^i_t) 
\end{align}
\end{subequations}
After the \ac{lstm} layer is a fully connected layer to benefit the \ac{lstm} layer output hidden state $h_t$. The full connected layer output is $h_{fc}$.
 \begin{equation}
h_{fc} = \sum_{i=1}^{N} w_{fc}^{j} \cdot h_t^{i} + b_{fc} 
\end{equation}
Where $N$ is the number of \ac{lstm} cells, $W_{fc}^i$ is the weight $b_{fc}$ is the bias for the  $c_i$ in the fully connected layer.
The trained weight and bias $Wy, by$ in the regression layer to predict the upcoming channel response value aiming to minimise the loss function $L$ between the \ac{lstm} predicted and true value $x_t$. 
\begin{subequations} 
\begin{align}
\tilde{H}(k,t+1) =  Wy.h_{fc}+by
\end{align}
 \begin{align}
 L(\tilde{H}(k,t+1),x_t)=\frac{1}{2}(\tilde{H}(k,t+1)-x_t)^2 \end{align}
\end{subequations} 
The normalised mean error \ac{nmse} in \ref{eq:nmse} is used to represent the prediction accuracy.
\begin{equation}
\label{eq:nmse}
\text{NMSE} = \frac{1}{n \cdot m} \sum_{t=1}^{n} \sum_{k=1}^{m} ( \frac{\tilde{H}_{dl}(k,t)-H_{dl}(k,t)}{\| {H_{dl}(k,t)} \|_2} )^2
\end{equation}
The normalised predicted $\tilde{H}_{dl}(k,t)$ is compared to the actual $H_{dl}(k,t)$ to show the accuracy of the prediction, where $n$ is the number of predictions and $m$ is the number of OFDM subcarrier. Thus, the NMSE is calculated for both the real and imaginary parts of the predicted channel gain, ultimately obtaining the average \ac{nmse}.    
\section{Simulated experimental scenario}\label{section3}

We analysed an urban microcell scenario involving a single transmitter and a mobile user with a fixed speed along a linear path. We used OFDM modulation after \ac{qam} modulation. The \ac{ssnr} is defined by $ \frac{ {E}\{P(k)^2\}}{\sigma^2_n} $ as shown in Equation \ref{eq:3b}. And the \ac{snr} is multiplied \ac{ssnr} by the channel gain $ |h|^2$. Decorrelation distance $d_{dec}$ is used to describe the similarity of channel variations, which depends on the distance between positions. The relative strength of the LoS component compared to the scattered components NLoS is stochastic variable $X^{kf}$.

\begin{figure}
\centering
\includegraphics[ width=0.5\textwidth]{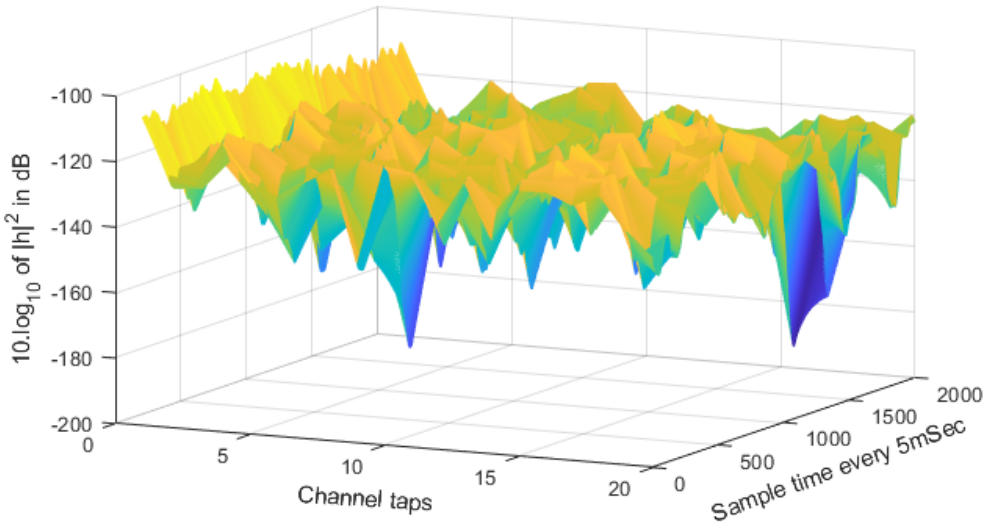}
\caption{Channel gain with 20 channel taps, mobile user in urban microcell scenario}
\label{fig:0}
\end{figure}

Figure \ref{fig:0} illustrates how signals between a mobile device and the transmitter are affected in the urban environment. The obstacles can block, reflect, or scatter the wireless signals, creating multiple paths (channel taps) for the signal to reach the mobile user. The time axis is included to show how the channel gain varies rapidly due to the movement of the mobile user or other dynamic factors in the urban environment.

To build a robust deep-learning LSTM model, we divided our dataset into an $80\%$ training set and a test set. We explored two approaches: an open-loop method based on previously estimated values and a closed-loop approach based on previously predicted values. The training and testing procedures were repeated several times with different generated channel state information, using the simulation setup described in table~\ref{tab:1}.

We evaluated the performance of the model by calculating the average of its predictions. The performance metric we used is the average of all predicted values for multiple subcarriers, as shown in the equation\ref{eq:nmse}.

This comprehensive analysis allowed us to assess the effectiveness and robustness of the model under multi-generated CSI under the same scenario conditions.

\begin{table}[htbp]
    \centering
    \caption{Simulation Setup Parameters}
    \begin{tabular}{|l|l|}
        \hline
        \textbf{Parameter} & \textbf{Value} \\
        \hline
        OFDM DFT size & 128 \\
        Number of Channel taps $L$ & 5 \\
        UE path-length & 100 m \\
        UE speed & 10 m/s \\
        Channel sampling time & 5 ms \\
        Decorrelation distance $d_{dec}$ & 5 m \\
       $X^{KF}$ parameter  & $\mu_{KF}=-3\si{dB}$, $\sigma_{KF}=0.5\si{dB}$ \\
        CSI size & 2000 \\
        Data set size & 20000 \\
        Modulation & 16QAM \\
        $\beta$ & 17/9 \cite{MMSEes} \\
        Training algorithm & adam \\
        Number of Hidden neurons $N$ &200\\
        \hline
    \end{tabular}
    \label{tab:1}
\end{table}
\section{Result}\label{section4}
After the estimation, we have two options to provide the \ac{dl} selector. Either we still 
interpolate the estimated channel frequency responses linearly or provide it to the channel predictor. The received pilot symbols in a frequency block manner offer the possibility of using an estimation method to estimate the channel's frequency response at each \ac{ofdm} sub-carrier. The \ac{ls} estimator provides a low-complexity method compared to the \ac{mmse} method due to the matrix inversion and the requirement to acknowledge the auto-correlation matrix of the channel responses.

\begin{figure}
\centering
\includegraphics[ width=0.45\textwidth]{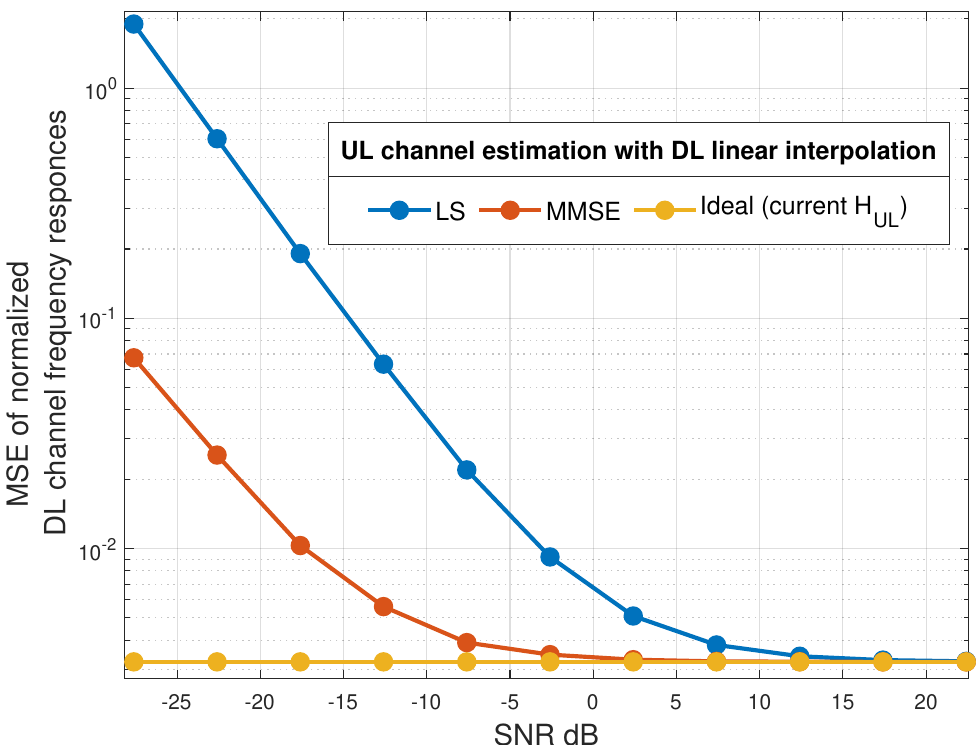}
\caption{DL channel information correctness in case of using linear interpolation with various estimation methods}
\label{fig:2}
\end{figure}

As illustrated in Figure \ref{fig:2}, the \ac{mmse} estimator performs better in bad channel conditions when the \ac{snr} is low, as it minimises the mean square error after linear interpolation in the \ac{dl} based on the pilot symbols via the \ac{ul} channel, in comparison to the \ac{ls} method. Both methods converge to the ideal estimation following interpolation under good channel conditions.

\begin{figure}
\centering
\includegraphics[ width=0.45\textwidth]{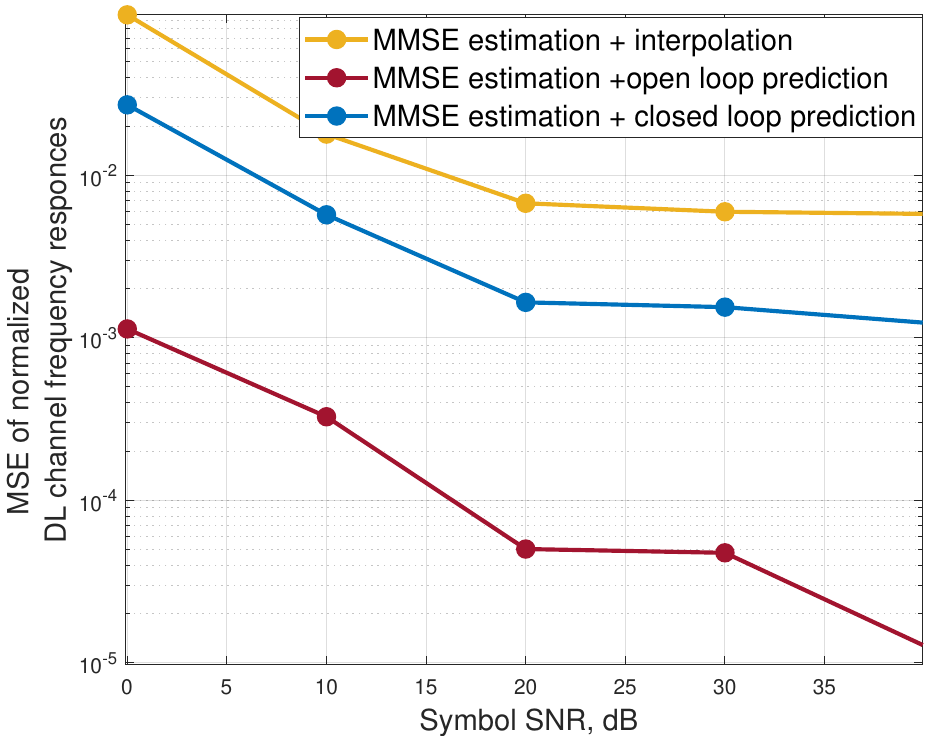}
\caption{DL channel information correctness in case of use prediction with high accurate estimation method}
\label{fig:3}
\end{figure}

Based on these results, we analyse the adaptive predictor, which has two modes for predicting the upcoming channel frequency responses, as shown in Figure \ref{fig:3}. Using the best estimation method, we compare the previous results in Figure \ref{fig:2} with our two proposed predictor modes. Despite the estimation error, the open-loop predictor provides highly accurate \ac{dl} channel frequency responses. This means that we should provide the predictor with highly accurate estimated \ac{ul} channel information each time slot.

Compared to the closed-loop predictor, which has lower accuracy than the open-loop, we do not need the estimated value for prediction but instead rely on the previously predicted values for the next output. This results in lower complexity compared to the open-loop predictor mode.
Therefore, when using \ac{mmse} estimation, there is a trade-off between complexity, computation, estimation frequency, and the accuracy of the input data for the link selector.

\begin{figure}
\centering
\includegraphics[ width=0.45\textwidth]{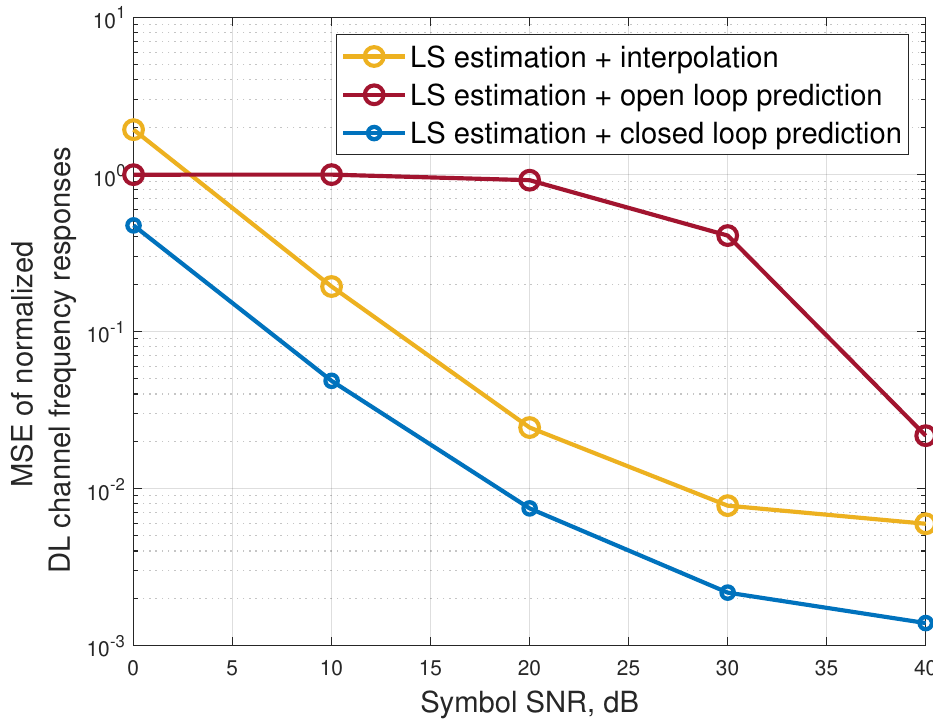}
\caption{DL channel information correctness in case of use prediction with low accurate estimation method}
\label{fig:4}
\end{figure}

In Figure \ref{fig:4}, we analyse the adaptive predictor in the case of a user with low-accuracy \ac{ls} estimation. Closed-loop prediction exhibits superior accuracy compared to the interpolation method. Therefore, closed-loop prediction mitigates complexity, particularly at low estimation, and enhances overall accuracy when applying LS estimation.
A significant challenge arises in the context of open-loop prediction utilising LS estimation due to the accumulation of estimation errors at the predicted output. This accumulation is a consequence of consecutive state updates within \ac{lstm} layers. Consequently, it leads to diminished accuracy in estimations and a marked reduction in prediction performance when contrasted with our baseline approach involving linear interpolation.

Hence, it is evident that both prediction modes hold the potential to generate output based on \ac{dl} selector input while considering long-term dependencies. However, the adaptability of the prediction system we propose is paramount. It enables the benefit of predictions based on the chosen estimation method and the specific accuracy requirements for the input data. This adaptability ensures the versatility of our system, enabling it to meet varied system complexity and deliver the desired level of precision. 

So, utilising closed-loop prediction results in heightened accuracy and reduced complexity, particularly at lower estimation frequencies, while open-loop prediction coupled with LS estimation may confront challenges associated with error accumulation. 
\section{Conclusion}\label{section5}
This paper presents a deep learning \ac{lstm} approach for predicting frequency-domain channel behaviour for a 3D stochastic fading channel within an urban environment, 3GPP geometric channel model. Our approach considers various factors, including user mobility, shadowing, and multipath effects.

The proposed method involves adaptive channel prediction based on pilot-aided estimation. We assess the accuracy of the predicted channel frequency responses, which serve as input to the 3D \ac{dl} selector decision-making process.

We proceed to analyse the prediction gain between two prediction possibilities. The first option involves making predictions after each estimation in an open-loop fashion, while the second entails employing consecutive closed-loop predictions for improved accuracy. This choice represents a trade-off between estimation requirements and prediction accuracy. In case estimation accuracy is high, consecutive closed-loop predictions can yield superior results. However, the closed-form prediction strategy may lose accuracy if the LS estimation accumulates estimation errors due to long-term dependencies.

The adaptive prediction strategy can balance estimation and prediction complexity and the required accuracy at the input of the 3D \ac{dl} selector.
 
\section*{Acknowledgements}
The authors acknowledge the financial support by the Federal Ministry of Education and Research of the Federal Republic of Germany (BMBF) in the project “Open6GHub” with funding number (DFKI-16KISK003K). The authors alone are responsible for the content of the paper.

\IEEEpeerreviewmaketitle
\bibliographystyle{IEEEtran}
\bibliography{Main.bib}

\begin{thebibliography}{10}
\providecommand{\url}[1]{#1}
\csname url@samestyle\endcsname
\providecommand{\newblock}{\relax}
\providecommand{\bibinfo}[2]{#2}
\providecommand{\BIBentrySTDinterwordspacing}{\spaceskip=0pt\relax}
\providecommand{\BIBentryALTinterwordstretchfactor}{4}
\providecommand{\BIBentryALTinterwordspacing}{\spaceskip=\fontdimen2\font plus
\BIBentryALTinterwordstretchfactor\fontdimen3\font minus \fontdimen4\font\relax}
\providecommand{\BIBforeignlanguage}[2]{{%
\expandafter\ifx\csname l@#1\endcsname\relax
\typeout{** WARNING: IEEEtran.bst: No hyphenation pattern has been}%
\typeout{** loaded for the language `#1'. Using the pattern for}%
\typeout{** the default language instead.}%
\else
\language=\csname l@#1\endcsname
\fi
#2}}
\providecommand{\BIBdecl}{\relax}
\BIBdecl

\bibitem{coexitenceuav}
M.~M. Azari, F.~Rosas, A.~Chiumento, and S.~Pollin, ``Coexistence of terrestrial and aerial users in cellular networks,'' in \emph{2017 IEEE Globecom Workshops (GC Wkshps)}, 2017, pp. 1--6.

\bibitem{UAVaidedcommunication}
Y.~Zeng, R.~Zhang, and T.~J. Lim, ``Wireless communications with unmanned aerial vehicles: opportunities and challenges,'' \emph{IEEE Communications Magazine}, vol.~54, no.~5, pp. 36--42, 2016.

\bibitem{5GWithUAVs}
M.~Mozaffari, A.~Taleb Zadeh~Kasgari, W.~Saad, M.~Bennis, and M.~Debbah, ``Beyond 5g with uavs: Foundations of a 3d wireless cellular network,'' \emph{IEEE Transactions on Wireless Communications}, vol.~18, no.~1, pp. 357--372, 2019.

\bibitem{hoverduration}
M.~Mozaffari, W.~Saad, M.~Bennis, and M.~Debbah, ``Wireless communication using unmanned aerial vehicles (uavs): Optimal transport theory for hover time optimization,'' \emph{IEEE Transactions on Wireless Communications}, vol.~16, no.~12, pp. 8052--8066, 2017.

\bibitem{adaptivetrasmissionairnode}
W.~Jiang and H.~D. Schotten, ``Predictive relay selection: A cooperative diversity scheme using deep learning,'' in \emph{2021 IEEE Wireless Communications and Networking Conference (WCNC)}, 2021, pp. 1--6.

\bibitem{pred1}
W.-S. Son and D.~S. Han, ``Analysis on the channel prediction accuracy of deep learning-based approach,'' in \emph{2021 International Conference on Artificial Intelligence in Information and Communication (ICAIIC)}, 2021, pp. 140--143.

\bibitem{pred2}
W.~Jiang and H.~D. Schotten, ``Recurrent neural network-based frequency-domain channel prediction for wideband communications,'' in \emph{2019 IEEE 89th Vehicular Technology Conference (VTC2019-Spring)}, 2019, pp. 1--6.

\bibitem{gain3dmodel}
S.~A. Busari, K.~M. Saidul~Huq, S.~Mumtaz, and J.~Rodriguez, ``Impact of 3d channel modeling for ultra-high speed beyond-5g networks,'' in \emph{2018 IEEE Globecom Workshops (GC Wkshps)}, 2018, pp. 1--6.

\bibitem{GBSM3D}
H.~Chang, J.~Bian, C.-X. Wang, Z.~Bai, J.~Sun, and X.~Gao, ``A 3d wideband geometry-based stochastic model for uav air-to-ground channels,'' \emph{IEEE Internet of Things Journal}, pp. 206--212, 2018.

\bibitem{GBSMin6G}
H.~Jiang, M.~Mukherjee, J.~Zhou, and J.~Lloret, ``Channel modeling and characteristics for 6g wireless communications,'' \emph{IEEE Network}, vol.~35, no.~1, pp. 296--303, 2021.

\bibitem{channelmodel2}
F.~Ademaj, S.~Schwarz, T.~Berisha, and M.~Rupp, ``A spatial consistency model for geometry-based stochastic channels,'' \emph{IEEE Access}, vol.~7, pp. 183\,414--183\,427, 2019.

\bibitem{3dsim}
M.~Kurras, S.~Dai, S.~Jaeckel, and L.~Thiele, ``Evaluation of the spatial consistency feature in the 3gpp geometry-based stochastic channel model,'' in \emph{2019 IEEE Wireless Communications and Networking Conference (WCNC)}, 2019, pp. 1--6.

\bibitem{3gpp}
3GPP, ``{5G; Study on channel model for frequencies},'' {3rd Generation Partnership Project (3GPP)}, Technical Specification (TS) 38.901, 2017, version 14.0.0.

\bibitem{estint}
S.~Coleri, M.~Ergen, A.~Puri, and A.~Bahai, ``Channel estimation techniques based on pilot arrangement in ofdm systems,'' \emph{IEEE Transactions on Broadcasting}, vol.~48, no.~3, pp. 223--229, 2002.

\bibitem{LSes}
S.~R, J.~T, A.~B. M, and R.~V, ``Channel estimation for ofdm systems using mmse and ls algorithms,'' in \emph{2022 6th International Conference on Trends in Electronics and Informatics (ICOEI)}, 2022, pp. 1--5.

\bibitem{MMSEes}
M.-H. Hsieh and C.-H. Wei, ``Channel estimation for ofdm systems based on comb-type pilot arrangement in frequency selective fading channels,'' \emph{IEEE Transactions on Consumer Electronics}, vol.~44, no.~1, pp. 217--225, 1998.

\end{thebibliography}
\end{document}